\documentclass[twocolumn,amsmath,amssymb,pre]{revtex4-1}

\usepackage{graphicx}
\usepackage{color}
\usepackage{dcolumn}
\usepackage{amsmath}
\usepackage{subfigure}
\usepackage{sidecap}

\usepackage{mathrsfs}
\usepackage{amsfonts}
\usepackage{indentfirst}
\usepackage{bm}
\usepackage{dsfont}
\usepackage[cmyk]{xcolor}

\usepackage[colorlinks,citecolor=blue]{hyperref}

\newcommand{\be}{\begin{equation}}
\newcommand{\ee}{\end{equation}}
\newcommand{\bey}{\begin{eqnarray}}
\newcommand{\eey}{\end{eqnarray}}
\newcommand{\bw}{\begin{widetext}}
\newcommand{\ew}{\end{widetext}}

\newcommand{\ra}{\rangle}
\newcommand{\la}{\langle}

\newcommand{\ba}{\begin{array}}
\newcommand{\ea}{\end{array}}
\newcommand{\bi}{\begin{itemize}}
\newcommand{\ei}{\end{itemize}}
\newcommand{\bem}{\begin{enumerate}}
\newcommand{\eem}{\end{enumerate}}

\begin{document}

\title{Statistics of phase space localization measures 
and quantum chaos in the kicked top model}

\author{Qian Wang$^{1,2}$ and Marko Robnik$^2$}

\affiliation{$^{1}$Department of Physics, Zhejiang Normal University, Jinhua 321004, 
People's Republic of China \\
$^2$CAMTP-Center for Applied Mathematics and Theoretical Physics, University of Maribor,
Mladinska 3, SI-2000, Maribor, Slovenia}

\date{\today}

\begin{abstract}

Quantum chaos plays a significant role in understanding
 several important questions of recent theoretical and experimental studies.
 Here, by focusing on the localization properties of eigenstates in phase space (by means of Husimi functions),
 we explore the characterizations of quantum chaos using the statistics of
 the localization measures, that is the inverse participation ratio and the Wehrl entropy.
 We consider the paradigmatic kicked top model, which shows a transition to chaos with increasing 
 the kicking strength.
 We demonstrate that the distributions of the localization measures exhibit a drastic change 
 as the system undergoes the crossover from integrability to chaos.
 We also show how to identify the signatures of quantum chaos from the central moments of 
 the distributions of localization measures. 
 Moreover, we find that the localization measures in the fully chaotic
 regime apparently universally exhibit the beta distribution, in agreement with previous
 studies in the billiard systems and the Dicke model.
 Our results contribute to a further understanding of quantum chaos and shed light on the
 usefulness of the statistics of phase space localization measures 
 in diagnosing the presence of quantum chaos, as well as the localization properties of eigenstates
 in quantum chaotic systems.

\end{abstract}

\maketitle

 \section{Introduction}

The importance of quantum chaos in understanding 
several fundamental questions that arise in recent experimental and theoretical works, 
has triggered a great deal of efforts in the study of quantum chaos in different areas of physics
\cite{Haake2010,Stockmann1999,Nandkishore2015,LucaD2016,Borgonovi2016,Deutsch2018,Maldacena2016,
Kos2018,Jahnke2019,Bertini2019,Lerose2020}.  
In contrast to the case of classical chaos, which is well defined as the exponential divergence of 
the closest orbits for initial perturbations
\cite{Lichtenberg2013,Cvitanovic2005}, 
the definition of quantum chaos in the narrow sense remains an open question \cite{Robnik2020}.
One, therefore, turns to focus on how to probe and measure the signatures 
of chaos in various quantum systems \cite{Haake2010,Stockmann1999,Magnani2016,ChanA2018,
Friedman2019,Mondal2020,Kobrin2021,Arias2021,Pausch2021,Fogarty2021}.
To date, the presence of chaos in a quantum system is commonly ascertained 
from its spectral properties.  
It is well known that the energy spectrum of quantum systems which are classically chaotic has universal
statistical properties, 
which are consistent with the predictions of the random matrix theory (RMT)
\cite{Casati1980,Mehta2004,BGS1984,Berry1985,Sieber2001,Guhr1998,Muller2004}.  
As a consequence, the RMT sets a benchmark to identify the emergence of chaos in quantum systems.
According to RMT, for example, a quantum system is said to be a chaotic system 
when its level spacing distribution follows the universal GOE/GUE/GSE level statistics, well approximated by the celebrated Wigner surmise \cite{Mehta2004,Wigner1958}, 
or its spectral form factor exhibits a robust linear ramp
\cite{Haake2010,Stockmann1999,Bertini2018,LiuJ2018,ChenX2018,Flack2020,PBraun2020,Khramtsov2021}.    

An alternative way to capture the onset of chaos in quantum systems is to investigate 
the structure of eigenstates.
The structure of eigenstates in quantum chaotic systems is essential to understand the thermalization
mechanism in isolated systems \cite{LucaD2016,Borgonovi2016,Deutsch2018,Deutsch1991,
Srednicki1994,Rigol2008}.
For quantum chaotic systems, it has been demonstrated that the mid-spectrum eigenstates are well 
described by the eigenstates of random matrices taken from the 
Gaussian ensembles of RMT \cite{Borgonovi2016,Kus1988}.
The eigenstates of Gaussian ensembles are random vectors with components that are independent
Gaussian random numbers.
Although this feature of eigenstates has been used as a witness of quantum chaos, several
remarkable exceptions in both single-particle
\cite{Fishman1982,Heller1984,Geisel1986,Casati1990,Bcker1997}
and many-body quantum chaotic systems \cite{Serbyn2021,Pilatowsky2021,Chandran2022}  
imply that further analysis on the structure of eigenstates in quantum chaotic systems is still required.

The structure of eigenstates can be examined in various ways, 
such as the fractality of the eigenstates \cite{Pausch2021,Backer2019,Tomasi2020,Pausch2021b}, 
the statistical properties of local observables in eigenstates
\cite{Rigol2008,Biroli2010,Rigol2012,Beugeling2014,Nakerst2021}, 
and the statistics of the eigenstate amplitudes \cite{Kus1988,Berry1977a,Brody1981,Zyczkowski1990,
Backer2007,Wouter2018,Srdinsek2021,Haque2022,LiRob1994}.  
In this work, we consider the localization characteristics of the quantum eigenstates. 
The eigenstates localization behavior for various quantum systems has been extensively explored 
in different contexts \cite{Pausch2021,Izrailev1990,Berkovits1998,Brown2008,
SantosF2010,Buccheri2011,SantosF2012,Serbyn2013,Beugeling2015,TorresH2017,DLuitz2020}.
To measure the degree of localization of an eigenstate, it is necessary to decompose it in some basis.
We use here the basis consisting of the coherent states. 
This means that we are interested in the phase space localization properties of the quantum eigenstates.
Coherent states, being the  states of minimal uncertainty,
and the derived Husimi functions are as close as possible to the
classical phase space structures, in particular in the semiclassical limit.
The phase space localization feature of quantum eigenstates has been explored in
kicked rotor \cite{Leboeuf1990b,Mirbach1995,Gorin1997,Izrailev1990,ManRob2013}, 
billiards \cite{Tomsovic1996,Mehlig1999,Cerruti2000,Batistic2013,Batistic2019,Lozej2022}, 
and Dicke model \cite{QWang2020,Villasenor2021,Pilatowsky2022}
in connection to the study of the localization phenomena observed in those systems. 
Here, by defining two different phase space localization measures, 
namely the inverse participation ratio and the Wehrl entropy,
we discuss how their statistical properties are affected by the onset of
chaos and show how the statistics of these measures 
tracks the transition between integrability and chaos in the kicked top model, one of
the paradigmatic models in the study of quantum chaos.

By decomposing a quantum eigenstate in the coherent state basis, we show that its phase space structure
is determined by the Husimi function \cite{Husimi1940} (squared modulus of the coherent states).
This represents quantum analogy of the
corresponding classical phase space, as a coherent state describing a phase space spot
is as close as possible to the classical phase space point.
As the system undergoes a transition from integrability to chaos, we observe a notable change in 
behaviors of the Husimi functions.
To quantitatively describe the structure of the system's eigenstates in phase space, we define two different
phase space localization measures in terms of the Husimi function.
We demostrate that the onset of chaos has strong impact on the 
distributions of the localization measures.
This leads us to show how to distinguish between integrability and chaos by means of the 
central moments of the distributions of localization measures.
We further show that the joint probability distribution of the localization measures also serves as a
diagnostic tool to signal the transition to quantum chaos.

 The article is organized as follows.
 In Sec.~\ref{secondS}, we introduce the kicked top model and briefly review 
 its chaotic features for both classical and quantum cases.
 Our detailed analysis on the statistical properties of the localization measures is presented
 in Sec.~\ref{thirdS}, where
 both the individual and joint statistics of the localization measures are investigated.
 We also discuss how to identify the signatures of quantum chaos 
 in the behaviors of central moments of the 
 localization measures distributions in this section.
 Finally, we conclude our study with a brief summary in Sec.~\ref{summary}.

 \section{Kicked top model}  \label{secondS}
 
 The system we have focused on in this work is the kicked top model, 
 which represents one of the prototypical models in studying 
 quantum chaos \cite{Haake2010} and has been realized in different 
 experimental platforms \cite{Chaudhury2009,Neill2016,Tomkovic2017}.
 Recently, we have studied it in the perspective of multifractal dimensions
 (entropies) of coherent states in the quasi-energy space \cite{WangRobnik2021}.
 These entropies describe the localization of the coherent states
 in the eigenbasis of the Floquet operator.

 The kicked top model describes a spin $\mathbf{J}=(J_x,J_y,J_z)$ evolving by
 the Hamiltonian (we set $\hbar=1$ throughout the work)\cite{Haake1987,Fox1994}
 \be \label{KTH}
    H=\alpha J_z+\frac{k}{2j}J_x^2\sum_{n=-\infty}^{+\infty}
             \delta(t-n),
 \ee
 where $\alpha$ denotes the precessional rotation angle of the spin around the $z$ axis
 between two kicks,
 $k$ is the strength of the (torsional) periodic kick, and 
 the components of $\mathbf{J}$ satisify the 
 commutation relations $[J_i,J_j]=i\epsilon_{ijk}J_k$.
 Here, the periodicity between kicks has been set equal to one. 
 We would like to point out that a different choice of $\alpha$ leads to
 the onset of chaos at different values of $k$ \cite{WangRobnik2021}.
 However, we have carefully checked that the main conclusions of this work are independent of the choice of $\alpha$. 
 Hence, we fixed $\alpha=4\pi/11$ throughout in our study.

 The time evolution of the system from kick to kick is governed by 
 the Floquet operator \cite{Haake1987}:
 \be \label{Foperator}
   \mathcal{F}=\exp\left(-i\frac{k}{2j}J_x^2\right)\exp\left(-i\alpha J_z\right).
 \ee
 The conservation of the magnitude $\mathbf{J}^2=j(j+1)$, due
 to the commutation with each $J_i$,
 leads us to express the Floquet operator
 in the basis consisting of Dicke states $\{|j,m\ra\}_{m=-j}^{m=j}$, which fulfill
 $\mathbf{J}^2|j,m\ra=j(j+1)|j,m\ra$ and $J_z|j,m\ra=m|j,m\ra$.
 Then, the matrix elements of $\mathcal{F}$ can be written as
 \be
    \la j,m|\mathcal{F}|j,m'\ra=\exp(-i\alpha m')D_{mm'},
 \ee
 where
 \begin{align}
    {D}_{mm'}&=\la j,m|\exp\left(-i\frac{k}{2j}J_x^2\right)|j,m'\ra, \notag \\
         &=\sum_{m_x=-j}^{m_x=j}\exp\left(-i\frac{k}{2j}m_x^2\right)\la j,m|j,m_x\ra\la j,m_x|j,m'\ra,
 \end{align}
 is the Winger ${D}$ function \cite{Rose1995} with $\{|j,m_x\ra\}_{m_x=-j}^{m_x=j}$ 
 being the eigenstates of $J_x$.
 The dimension of the matrix is given by $\mathcal{D}_\mathcal{H}=2j+1$. 
 However, since $\mathcal{F}$ commutes with the parity operator
 $\Pi=e^{i\pi(j+J_z)}$, the matrix can be decomposed into an even-parity block
 with dimension $\mathcal{D}_e=j+1$ and the other odd-parity with $\mathcal{D}_o=j$.
 In this work, we only consider the even-parity subspace.

 In the Heisenberg picture, the spin operators are evolved by the Heisenberg equation \cite{Fox1994}: 
 \be \label{QMJ}
     \mathbf{J}(n+1)=\mathcal{F}^\dag\mathbf{J}(n)\mathcal{F}.
 \ee 
 By using the commutation relations between spin operators and the Campbell identity
 \be \label{Campbell}
    e^{\eta A}Be^{-\eta A}=B+\eta[A,B]+\frac{\eta^2}{2!}[A,[A,B]]+\ldots,
 \ee
 the explict form of the Heisenberg equation can be written as
 \begin{align}\label{QEJ}
    J_x(n+1)&=J_x(n)\cos\alpha-J_y(n)\sin\alpha,  \notag \\
    J_y(n+1)&=\frac{1}{2}\Theta_n
        \exp\left[i\frac{k}{2j}\Upsilon_n\right]+\mathrm{H.c.}, 
         \notag \\
    J_z(n+1)&=\frac{1}{2i}\Theta_n
        \exp\left[i\frac{k}{2j}\Upsilon_n\right]+\mathrm{H.c.}, 
 \end{align}
 where $\Theta_n=\left[J_x(n)\sin\alpha+J_y(n)\cos\alpha+iJ_z(n)\right]$ and 
 $\Upsilon_n=2[J_x(n)\cos\alpha-J_y(n)\sin\alpha]+1$. 
 In the Appendix we present a short derivation of the above formulas.
 With increasing the kicking strength $k$, the model undergoes a transition from integrability to chaos.
 To see this, we will first analyze the emergence of chaos in the classical kicked top model, which can
 be obtained by taking the classical limit of the Heisenberg equation in (\ref{QEJ}).
 Then we show how the chaos manifests itself in the quantum kicked top model through the
 spectral statistics of the Floquet operator (\ref{Foperator}).

  \begin{figure}
  \includegraphics[width=\columnwidth]{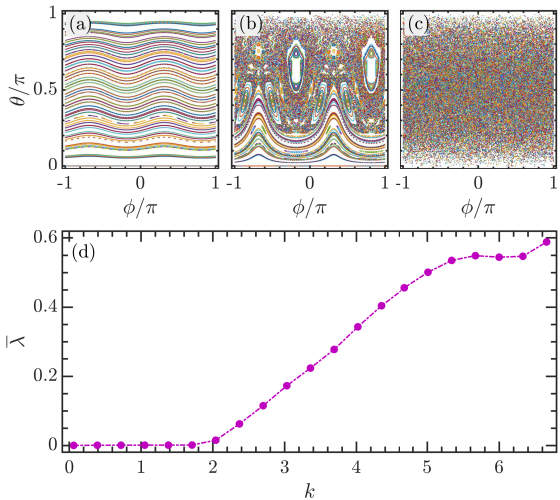}
  \caption{(a)-(c): Phase space portraits of the classical kicked top model for $k=0.4$ (a),
  $k=2.4$ (b), and $k=6$ (c). 
  The variables ($\theta, \phi$) are plotted for $300$ initial conditions, 
  each with a duration of $300$ kicks.
   (d) Phase space averaged Lyapunov exponent, $\bar{\lambda}$, 
   as a function of the kick strength $k$.
   In a numerical simulation, $\bar{\lambda}$ is calculated by averaging over $40000$ trajectories, 
   each evolved for $4000$ kicks.
   Other parameter: $\alpha=4\pi/11$.}
  \label{PLyE}
 \end{figure}

\subsection{Classical kicked top model} \label{SecSa}

The classical counterpart of the Heisenberg equation in (\ref{QEJ}) 
is obtained in the limit $j\to\infty$.
To see this, we define the scaled vector $\mathbf{X}=\mathbf{J}/j$ which obeys the
commutation relations $[X_a, X_b]=(1/j)i\epsilon_{abc}X_c$. 
As $j\to\infty$, the vanishing of commutators between the components of $\mathbf{X}$ implies
that $X_a$ become classical variables.
Then, substituting $\mathbf{X}$ into Eq.~(\ref{QEJ}), after some algebra, we find the classical map
can be written as
\begin{align} \label{CEM}
\begin{bmatrix}
  X_{n+1}  \\
  Y_{n+1}  \\
  Z_{n+1}
 \end{bmatrix}
  =
  \begin{bmatrix}
  \cos\alpha  &   -\sin\alpha  &   0   \\
  \sin\alpha\cos\Omega_n  &   \cos\alpha\cos\Omega_n   &   -\sin\Omega_n  \\
  \sin\alpha\sin\Omega_n   &   \cos\alpha\sin\Omega_n    &  \cos\Omega_n
  \end{bmatrix}
   \begin{bmatrix}
   X_n  \\
   Y_n  \\
   Z_n
  \end{bmatrix},
\end{align}
where $\Omega_n=k(X_n\cos\alpha-Y_n\sin\alpha)$.
The conservation of $\mathbf{J}^2$ entails $|\mathbf{X}|^2=X^2+Y^2+Z^2=1$.
This means that the classical variables $(X,Y,Z)$ lie on the unit sphere and
they can be parametrized in terms of the azimuthal angle $\theta$ and polar angle $\phi$
as follows: $X=\sin\theta\cos\phi$, $Y=\sin\theta\sin\phi$, and $Z=\cos\theta$.
Hence, the classical phase space is actually two dimensional space described by variables
$\theta=\cos^{-1}Z$ and $\phi=\tan^{-1}(Y/X)$.

A prominent feature exhibited by the kicked top model is the transition to chaos 
as the kicking strength $k$ increases.
It is known that the model shows regular behavior in the phase space for lower values of $k$,
while increasing $k$ gives rise to the extension of the chaotic regime in the phase space and,
therefore, increases the degree of chaos \cite{Haake1987}.  
The emergence of chaos in the dynamics of the classical kicked top with increasing $k$ 
is clearly observed in Figs.~\ref{PLyE}(a)-\ref{PLyE}(c), where the Poincar\'e sections for
several values of $k$ are plotted.
One can see that the phase space turns from regular motion for small $k$ to the 
mixed dynamics with regular regions embedded in the chaotic sea for larger $k$.
For even larger $k$, as the $k=6$ case plotted in Fig.~\ref{PLyE}(c), the phase space is governed
by globally chaotic dynamics.

To quantitatively analyze the chaotic properties of the model, we consider the phase space averaged 
Lyapunov exponent, which measures the degree of chaos in the model and is defined as
\be
   \bar{\lambda}= \frac{1}{4\pi}\int dS\lambda_m,
\ee  
where $dS=\sin\theta d\theta d\phi$ is the phase space area element 
(or Haar measure) \cite{Ariano1992}
and $\lambda_m$ denotes the largest Lyapunov exponent of the classical map in (\ref{CEM}). 
The largest Lyapunov exponent quantifies the rate of deviation between two nearby orbits
in a dynamical system \cite{Vallejos2002,Jayaraman2006}.
For the kicked top model, it can be calculated as \cite{Ariano1992,Constantoudis1997,Parker2012}
\be
   \lambda_m=\ln\left[\lim_{n\to\infty}(t_m)^{1/n}\right],
\ee
where $t_m$ represents the largest eigenvalue of $T=\prod_{p=1}^n\mathfrak{T}(\mathbf{X}_p)$
and $\mathfrak{T}(\mathbf{X}_n)=\partial\mathbf{X}_{n+1}/\partial\mathbf{X}_n$
is the tangent map of Eq.~(\ref{CEM}).

By averaging the largest Lyapunov exponents over different initial conditions in the phase space
for various values of $k$, 
we show the dependence of $\bar{\lambda}$ on the kicking strength $k$ in Fig.~\ref{PLyE}(d).
The regularity of the model at small $k$ leads to the zero value of $\bar{\lambda}$ and
 it keeps zero up to a certain kicking strength $k_c\approx2$,  from which it starts to grow with  
 increasing $k$. 
 Hence, the model undergoes a transition from integrability to chaos as $k$ is increased 
 and the level of chaos is enhanced by increasing the kicking strength.

  \begin{figure}
   \includegraphics[width=\columnwidth]{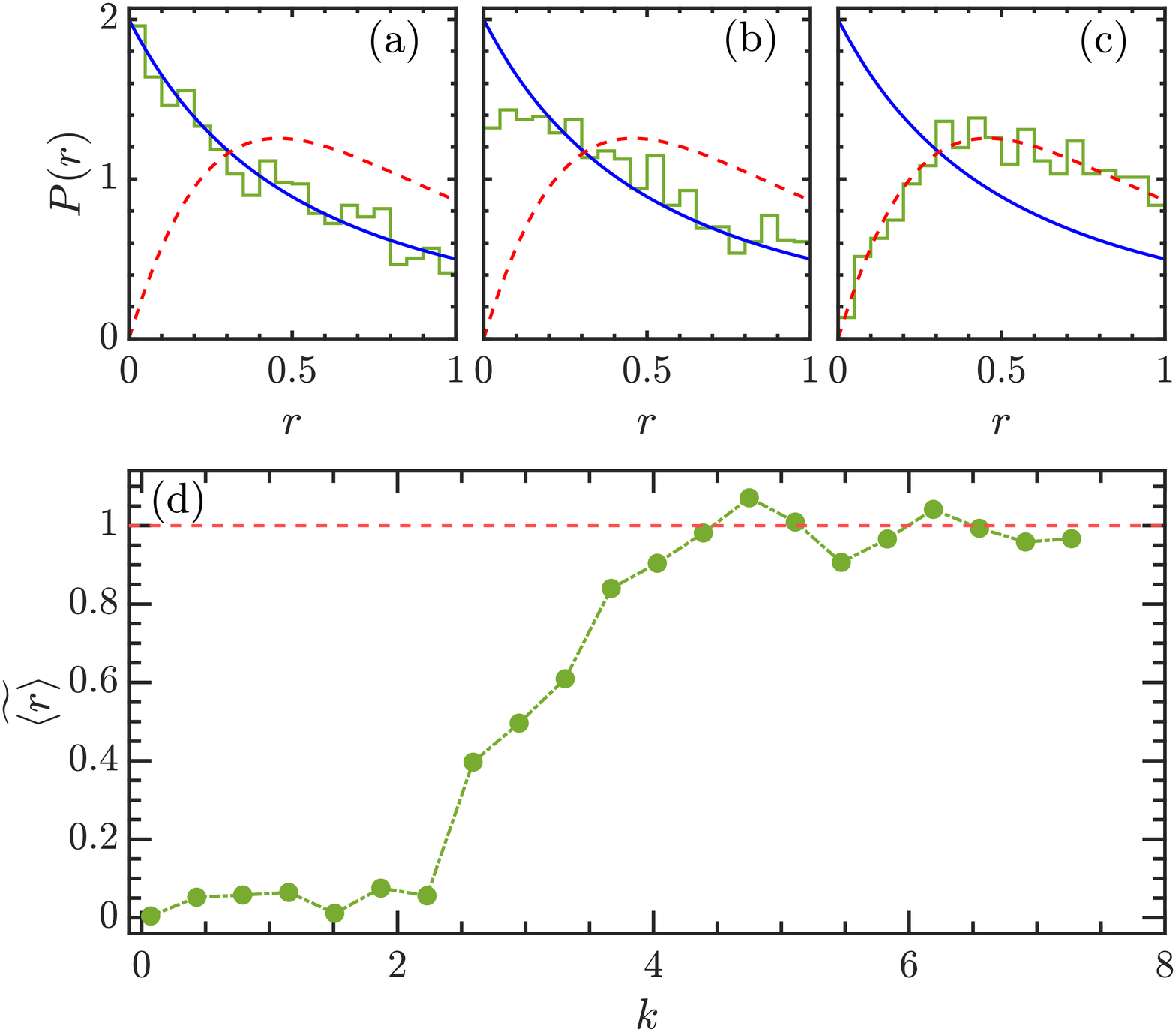}
   \caption{(a)-(c) Distribution of the consecutive level spacing ratios $P(r)$ for
   $k=0.4$ (a), $k=2.4$ (b), and $k=6$ (c).
   In each panel, the blue solid and red dashed curves denote the $P_{\mathrm{P}}(r)$ and 
   $P_{\mathrm{WD}}(r)$ [cf.~Eq.~(\ref{AnPr})], respectively.
   (d) Rescaled average ratio $\widetilde{\la r\ra}$ in (\ref{Avrn}) as a function 
   of the kicking strength $k$.
   The red dashed horizontal line represents $\widetilde{\la r\ra}=1$.
   Other parameters: $j=2000$ and $\alpha=4\pi/11$.}
   \label{Prvsk}
 \end{figure}

\subsection{Chaos in quantum kicked top model} \label{SecSb}

The chaotic properties discussed above in the classical kicked top model are also
manifested in its quantum counterpart.
The signatures of quantum chaos can be captured by several probes,
including the statistics of eigenvalues and eigenvectors \cite{Brody1981,Zyczkowski1990},
the dynamical behavior of the Loschmidt echo \cite{Emerson2002,Torres2017},
the entanglement dyanmics \cite{WangX2004,Gietka2019,Piga2019}, 
the out-of-time-ordered correlators \cite{Mata2018,Fortes2019,Lantagne2020}, 
quantum coherence \cite{Anand2021},
and the operator complexity \cite{Magan2018,Bhattacharyya2021,Rabinovici2022}, to name a few.
Here, in order to unveil the fingerprint of chaos in the quantum kicked top model, we
study the spectral properties of the Floquet operator based on the level spacing ratios $r_n$,
defined as \cite{Oganesyan2007,Atas2013}
\be
r_n=\frac{\mathrm{min}(d_n,d_{n+1})}{\mathrm{max}(d_n,d_{n+1})}
=  {\mathrm{min}(w_n,\frac{1}{w_n})} ,
\ee   
where $d_n=\mu_{n+1}-\mu_n$  is the spacing between two successive levels 
with $\mu_n$ being the $n$th eigenphase of the Floquet operator,
and $w_n=d_{n+1}/d_n$.
Clearly, the definition of $r_n$ implies that it varies in the interval $r_n\in[0,1]$.
As $r_n$ does not rely on the local density of states, 
the calculation of its distribution is not required to perform the so-called unfolding procedure,
which is known to be intricate \cite{Gomez2002,Corps2021}.    
This makes the level spacing ratio $r_n$ a 
convenient chaos indicator in the studies of quantum chaos, 
in particular for quantum many-body chaotic systems.

It has been demonstrated that the distribution of $r_n$, denoted by $P(r)$, 
can be used to distinguish between integrable and chaotic systems \cite{Oganesyan2007,Atas2013}.
In particular, the analytical expression of $P(r)$ has been obtained for both 
integrable (Poisson statistics) and chaotic (Wigner-Dyson statistics) spectra, 
given by \cite{Atas2013,Atas2013b,Giraud2022}
\be \label{AnPr}
   P_{\mathrm{P}}(r)=\frac{2}{(1+r)^2},\ 
   P_{\mathrm{WD}}(r)=\frac{27}{4}\frac{r+r^2}{(1+r+r^2)^{5/2}}.
\ee
The distributions of level spacing ratios $P(r)$ of the Floquet operator for several values of $k$
are shown in Figs.~\ref{Prvsk}(a)-\ref{Prvsk}(c), where we also compare our numerical results
to the analytical formula of $P(r)$ in Eq.~(\ref{AnPr}).
We see that $P(r)$ follows the distribution for Poisson statistics $P_\mathrm{P}(r)$ for small $k$
[Fig.~\ref{Prvsk}(a)].
This means the absence of level repulsions in the model are consistent with the 
regular structure of the phase space.
As $k$ is increased, $P(r)$ deviates from $P_\mathrm{P}(r)$ and has
a Poisson-like tail, as observed in Fig.~\ref{Prvsk}(b).
This indicates the highly localized weak correlations between eigenphases corresponding to
the mixed feature in the phase space.
Finally, for the case of strong kicking strength, as shown in Fig.~\ref{Prvsk}(c), the ratio distribution
$P(r)$ is in good agreement with $P_\mathrm{WD}(r)$ suggesting level repulsion 
in the fully chaotic regime.
The changes in the behaviors of the distribution $P(r)$ clearly confirms the 
integrablity-to-chaos transition in the kicked top model.
It is worthwhile to mention that the level statistics for the Floquet operator 
in the chaotic regime should belong to random matrices of the circular orthogonal ensemble (COE).
However, since the COE statistics is asymptotically described by the 
random matrices belonging to the Gaussian orthogonal ensemble (GOE) 
in the thermodynamic (also semiclassical) limit \cite{Alessio2014}, we have therefore 
compared our numerical results to the GOE counterpart.

 \begin{figure*}
  \includegraphics[width=\textwidth]{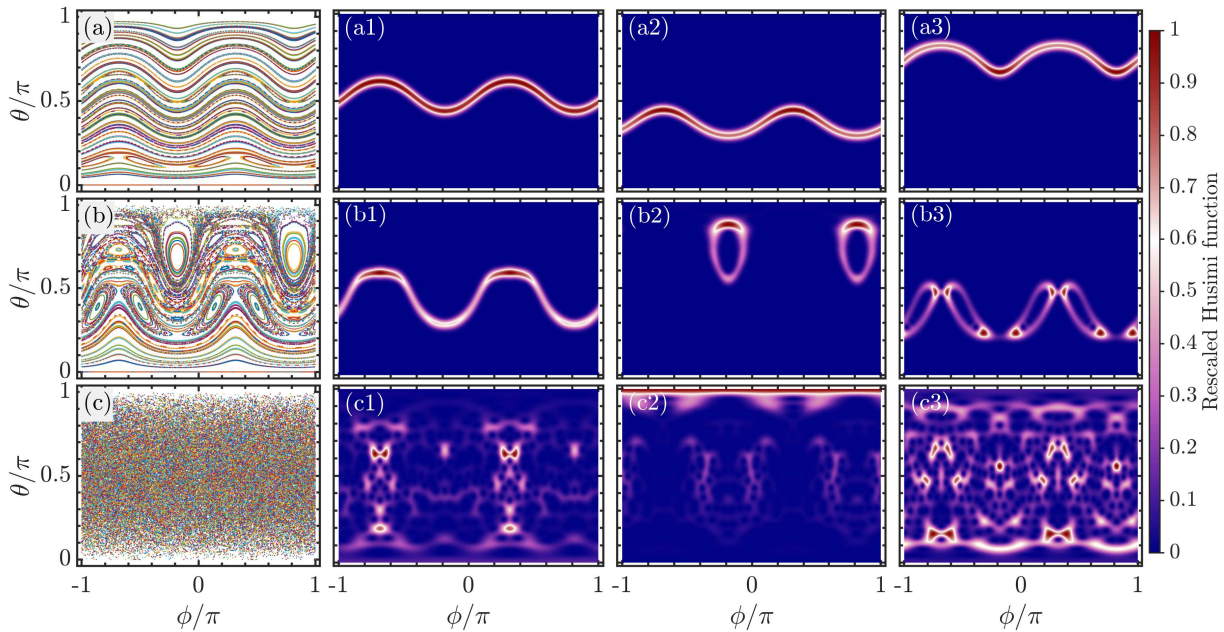}
  \caption{(a)-(c): Phase portraits of the classical kicked top model for 
  (a) $k=1$, (b) $k=2$, and (c) $k=7$. 
  Here, $300$ randomly chosen initial conditions in the phase space 
  have been plotted after $300$ kicks.
  (a1)-(a3): Husimi function rescaled by its maximum value for the eigenstates of $\mathcal{F}$ with
  eigenphases (a1) $\mu\approx-2.9 $, (a2) $\mu\approx-2.19$, and (a3) $\mu\approx2.17$
  for $k=1$.
  (b1)-b(3): Husimi function rescaled by its maximum value for the 
  eigenstates of $\mathcal{F}$ with same eigenphases as in
  (a1)-a(3) for $k=2$.
  (c1)-c(3): Husimi function rescaled by its maximum value for the eigenstates of $\mathcal{F}$ 
  with same eigenphases as in (a1)-a(3) for $k=7$.
   Other parameters are: $\alpha=4\pi/11$ and $j=150$.}
  \label{Hfunction}
 \end{figure*}

Instead of focusing on the ratio distribution $P(r)$, the crossover from integrable to chaos in the model
can also be captured by the mean level spacing ratio $\la r\ra$, defined as
\be
  \la r\ra=\frac{1}{\mathcal{N}}\sum_{n=1}^\mathcal{N}r_n,
\ee
where $\mathcal{N}$ denotes the total number of $r_n$.
For integrable systems, the Poisson statistics yields $\la r\ra_P\approx0.39$ \cite{Atas2013},
while for the chaotic systems with Wigner-Dyson statistics, the mean value 
$\la r\ra_{WD}\approx0.53$ \cite{Atas2013,Giraud2022,Alessio2014}.
It is more convenient to consider the rescaled average level spacing ratio \cite{Patrycja2022}:
\be \label{Avrn}
  \widetilde{\la r\ra}=\frac{\left|\la r\ra-\la r\ra_P\right|}{\la r\ra_{WD}-\la r\ra_P}.
\ee
It is defined in the range $0\leq\widetilde{\la r\ra}\leq1$, with two limiting values corresponding
to the Poisson and Wigner-Dyson distributions, respectively.
Figure~\ref{Prvsk}(d) illustrates how $\widetilde{\la r\ra}$ varies as a 
function of the kicking strength $k$.
The transition to chaos with increasing $k$ is evidently revealed by the interpolation of 
$\widetilde{\la r\ra}$ between Poisson and Wigner-Dyson cases.   
We further note that the upturn in $\widetilde{\la r\ra}$ with increasing $k$ is in agreement
with that of the phase space averaged Lyapunov exponent in Fig.~\ref{PLyE}(d),
indicating a good quantum-classical correspondence.

\section{Statistics of the phase space localization measures} \label{thirdS}

The transition to chaos also correlates with a remarkable change in the structure of eigenstates.
To analyze the structure of eigenstates, it is necessary to decompose them in a certain basis.
The choice of basis is usually determined by the system and the physical question under consideration.
In this work we aim to explore the properties of the phase space localization. 
A natural choice for us is the basis consisting of the coherent states $|\theta,\phi\ra$,
as they are as close as possible to the classical phase points due to their minimal uncertainty.
For the kicked top model, the coherent states are the generalized $\mathrm{SU}(2)$ 
spin coherent states, which are generated by rotating the Dicke state $|j,j\ra$ 
as follows \cite{Gazeau2009,ZhangW1990,Antoine2018}:
\begin{align}
   |\theta,\phi\ra&=\exp[i\theta(J_x\sin\phi-J_y\cos\phi)]|j,j\ra, \notag \\
      &=\sum_{m=-j}^{m=j}\frac{\tau^{j-m}}{(1+|\tau|^2)^j}\sqrt{\frac{(2j)!}{(j-m)!(j+m)!}}|j,m\ra,
\end{align}
with $\tau=\tan(\theta/2)e^{i\phi}$ and $\theta\in[0,\pi), \phi\in[-\pi,\pi)$.
The coherent states basis $\{|\theta,\phi\ra\}$ is overcomplete with the closure relation
\be
   \mathds{1}=\frac{2j+1}{4\pi}\int d\theta d\phi\sin\theta|\theta,\phi\ra\la\theta,\phi|.
\ee
The expansion of the $n$th eigenstate $|\mu_n\ra$ of $\mathcal{F}$ 
in the basis $\{|\theta,\phi\ra\}$ can be written as
\be
   |\mu_n\ra=\frac{2j+1}{4\pi}\int d\theta d\phi\sin\theta p_n(\theta,\phi)|\theta,\phi\ra,
\ee
where $p_n(\theta,\phi)=\la\theta,\phi|\mu_n\ra$ is the overlap between the coherent state 
$|\theta,\phi\ra$ and the $n$th eigenstate $|\mu_n\ra$.
Then, valuable information about the structure of the $n$th eigenstate in the phase space 
is provided by the Husimi function, defined as the square of $p_n(\theta,\phi)$ module,
\be
   Q_n(\theta,\phi)=|p_n(\theta,\phi)|^2=\la\theta,\phi|\rho_n|\theta,\phi\ra,
\ee
with $\rho_n=|\mu_n\ra\la\mu_n|$ and it satisfies the normalization condition
\be
  \frac{2j+1}{4\pi}\int d\theta d\phi\sin\theta Q_n(\theta,\phi)=1.
\ee

 \begin{figure}
  \includegraphics[width=\columnwidth]{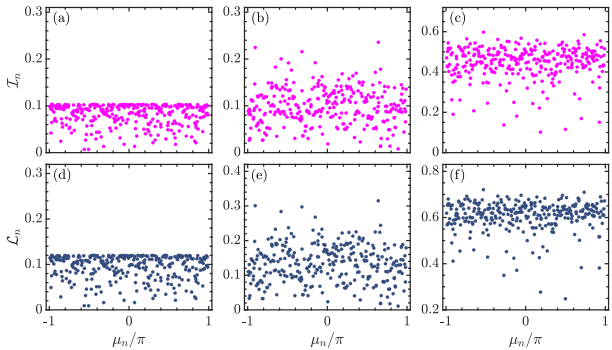}
  \caption{(a)-(c): Inverse participation ratio $\mathcal{I}_n$ of the eigenstates for the kicked top model
  as a function of eigenphases $\mu_n$ of the Floquet operator for 
  (a) $k=0.4$, (b) $k=2.4$, and (c) $k=6$. 
  (d)-(f): Localization measure $\mathcal{L}_n$ for the eigenstates of the kicked top model
  as a function of eigenphases $\mu_n$ of the Floquet operator for the same values of $k$ as in
  panels (a)-(c).
  Other parameters: $\alpha=4\pi/11$ and $j=150$.}
  \label{LMSEg}
 \end{figure}

The Husimi functions of various eigenstates of $\mathcal{F}$ for several values of $k$ 
are plotted in Fig.~\ref{Hfunction} with associated classical phase portraits, in agreement
with the principle of uniform semiclassical condensation (PUSC) of the Wigner functions or
Husimi functions - see \cite{Robnik1998,Lozej2022,VRL1999} and references therein. (Husimi
function also is a Gaussian smoothed Wigner function.)
We first note that the Husimi functions show a good correspondence to 
the classical phase space orbits.   
This is due to the fact that the coherent states represent the closest quantum analog 
of the classical phase space points.
Meanwhile, the good agreement between the Husimi function and the classical phase space structure also confirms that the coherent state basis is an appropriate basis for investigating 
the phase space localization. 
We further observe the degree of localization of the Husimi function in the phase space
depending on the kicking strength. 
In particular, even in deep chaotic regime, there still exist some eigenstates that are highly localized
in the phase space, such as the one illustrated in Fig.~\ref{Hfunction}(c2).

 \begin{figure}
  \includegraphics[width=\columnwidth]{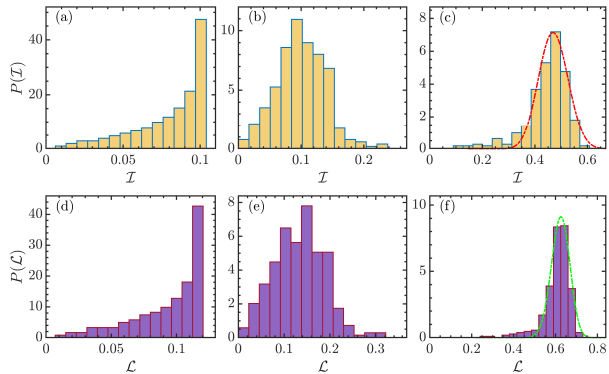}
  \caption{(a)-(c): Distribution of $\mathcal{I}_n$ for (a) $k=0.4$, (b) $k=2.4$, and (c) $k=6$. 
  In panel (c), the red dot-dashed line denotes the beta distribution (\ref{BtDs}) with
  $(x_{min}, x_{max})=(0.145, 0.872)$ and the shape parameters are $a=18.9435$ and $b=23.3284$.
  (d)-(f): Distribution of $\mathcal{L}_n$ for the same values of $k$ as in panels (a)-(c).
  The green dot-dashed line in panel (f) represents the beta distribution (\ref{BtDs}) with
  $(x_{min}, x_{max})=(0.219, 0.8849)$ and
  the shape parameters $(a,b)=(33.7363, 21.8592)$.
   Other parameter: $\alpha=4\pi/11$ and $j=150$.}
  \label{PdfLMS}
 \end{figure}

To measure the degree of localization of the $n$th eigenstate of $\mathcal{F}$ 
in the basis $\{|\theta,\phi\ra\}$, we consider two different 
localization measures based on the Husimi function.
The first one is the well-known inverse participation ratio, which measures 
how many basis states the quantum state occupies \cite{Ever2008}.
In terms of Husimi function, the inverse participation ratio for the $n$th eigenstate is defined as
\be
  \mathcal{I}_n=
  \left[\mathcal{N}_c\frac{2j+1}{4\pi}\int d\theta d\phi\sin\theta Q_n^2(\theta,\phi)\right]^{-1},
\ee 
where $\mathcal{N}_c=(2j+1)/(4\pi)\int d\theta d\phi\sin\theta=2j+1$ acts as 
the normalization constant.
The second localization measure is defined through the Wehrl entropy \cite{Wehrl1978} and 
for the $n$th eigenstate it is given by \cite{Batistic2013,QWang2020,Lozej2022}
\be
   \mathcal{L}_n=\frac{\exp(S_W)}{\mathcal{N}_c},
\ee
where 
\be
  S_W=-\frac{2j+1}{4\pi}\int d\theta d\phi\sin\theta Q_n(\theta,\phi)\ln[Q(\theta,\phi)],
\ee 
is the Wehrl entropy. 
In the semiclassical limit with $j\to\infty$, both localization measures 
interpolate between two extreme ends: the uttermost
localized eigenstates with $\mathcal{I}_n=\mathcal{L}_n=0$ 
and the fully delocalized chaotic eigenstates with $\mathcal{I}_n=\mathcal{L}_n=1$
\cite{QWang2021a,Lozej2022}. In the former case $Q_n$ is localized on a small
region of size $\Delta S$ and is constant there,
so that $\mathcal{I}_n$ and $\mathcal{L}_n$ go to zero as
 $\Delta S$ tends to zero.
In the latter case $Q_n$ is constant and equal to its average value
 $\bar{Q}_n= 1/(2j+1)$. 
 
 \begin{figure*}
  \includegraphics[width=\textwidth]{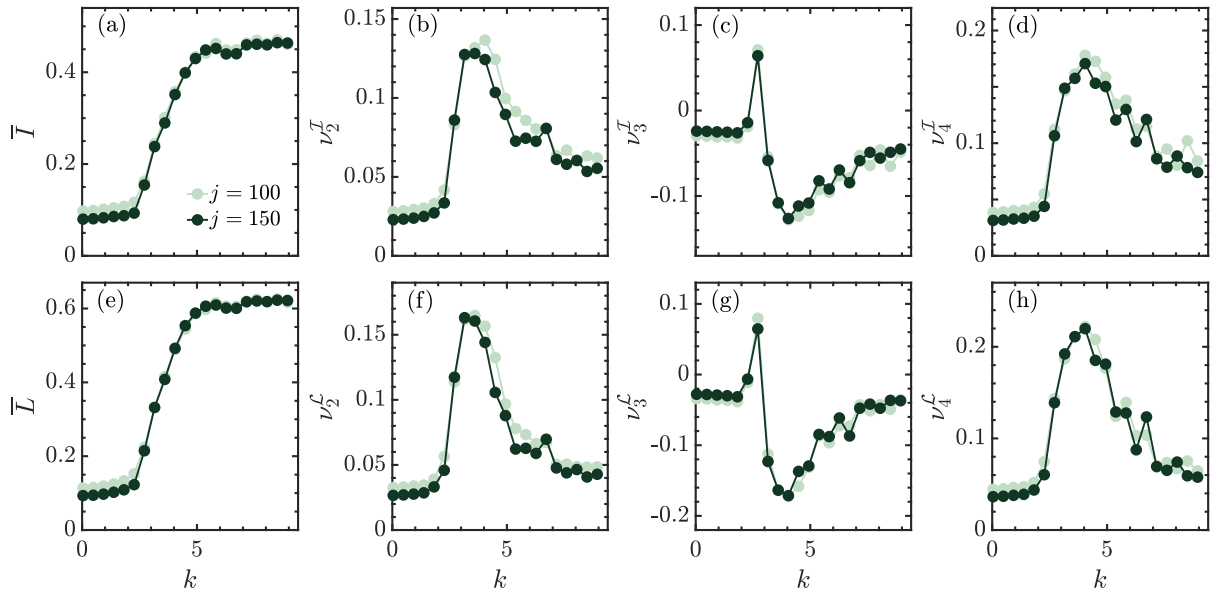}
  \caption{(a) Mean value $\bar{I}$, 
  (b) Standard deviation $\nu_2^{\mathcal{I}}$,
  (c) Cube root of the third central moment $\nu_3^{\mathcal{I}}$,
  and (d) $\nu_4^\mathcal{I}$, obtained from Eq.~(\ref{RootMs}),
  as a function of $k$ for several system sizes, see the legend in panel (a). 
  Dependence on $k$ of (e) mean value $\bar{L}$, (f) standard deviation $\nu_2^{\mathcal{L}}$,
  (g) cube root of the third central moment $\nu_3^{\mathcal{L}}$, and
  (h) $\nu_4^\mathcal{L}$, calculated by Eq.~(\ref{RootMs}), 
  for the same system sizes as legend in panel(a).
  Other parameter: $\alpha=4\pi/11$.}
  \label{LMScms}
 \end{figure*}

In the following of this section, we will focus on both the individual and joint statistics 
of these localization measures. 
The purpose of our study is to unveil the impact of chaos on the structure of eigenstates and 
to identify the signatures of chaos in the statistical properties of the phase space localization measures.

\subsection{Statistics of $\mathcal{I}_n$ and $\mathcal{L}_n$}

Let us consider the statistics of the localization measures $\mathcal{I}_n$ and $\mathcal{L}_n$.
In Fig.~\ref{LMSEg}, we plot $\mathcal{I}_n$ and $\mathcal{L}_n$ 
as a function of $\mu_n$ for different values of $k$.
As both $\mathcal{I}_n$ and $\mathcal{L}_n$ measure the degree of 
(de)localization of an eigenstate in the phase space, one can expect that 
they should behave in a similar way as a function of eigenphases.
This is confirmed by our numerical results which show overall 
similarities between the behaviors of $\mathcal{L}_n$ and $\mathcal{I}_n$.
We see that the values of $\mathcal{I}_n$ and $\mathcal{L}_n$ are low
and concentrated in a narrow range for small $k$ [see Figs.~\ref{LMSEg}(a) and \ref{LMSEg}(d)], 
suggesting that the eigenstates are localized in the phase space reflecting the regular 
dynamics in the classical case.
Moreover, there is an obvious concentration in $\mathcal{I}_n$ and $\mathcal{L}_n$ around their
corrresponding maximal values.
A careful check shows that these sharp upper limits of the localization measures 
can only be seen in the regular regime and 
are associated with the eigenstates that correspond to the 
classical orbits located in an interval with $\pi/4\lesssim\theta\lesssim3\pi/4$. 
It should be observed that the thickness of the Husimi function localized on an
invariant torus of length $\approx 2\pi$ is of the order of the square root
of the effective Planck constant $\hbar_{eff} \approx 1/j$, which determines
the maximum value of $\mathcal{I}_n$ and $\mathcal{L}_n$. This is of course
in contradistinction of the chaotic eigenstates.

As $k$ is increased, the values of localization measures also increase, 
but the eigenstates with high and low values of the localization measures are coexisting in the spectrum, 
as illustrated in Figs.~\ref{LMSEg}(b) and \ref{LMSEg}(e).
This means that the degree of localization has strong fluctuations among the eigenstates. 
The reason can be attributed to the mixed feature exhibited by the corresponding classical dynamics,
in agreement with PUSC.
The eigenstates associated with the regular islands are highly localized states 
with low-$\mathcal{I}_n/\mathcal{L}_n$, while the high-$\mathcal{I}_n/\mathcal{L}_n$ values 
stem from the eigenstates that are located in the chaotic sea.
At large values of $k$, as plotted in Figs.~\ref{LMSEg}(c) and \ref{LMSEg}(f), 
the values of $\mathcal{I}_n$ and $\mathcal{L}_n$ are much larger.
This is due to the fact that the system becomes classically globally chaotic, which leads to
the eigenstates being spread over the whole phase space.
However, there are still some low-$\mathcal{I}_n/\mathcal{L}_n$ eigenstates, 
which are the localized chaotic eigenstates, as observed in Fig.~\ref{Hfunction}(c2). 
For a phenomenological analysis
of Husimi functions in a mixed-type regime see \cite{Lozej2022} and references therein.

 \begin{figure*}
  \includegraphics[width=\textwidth]{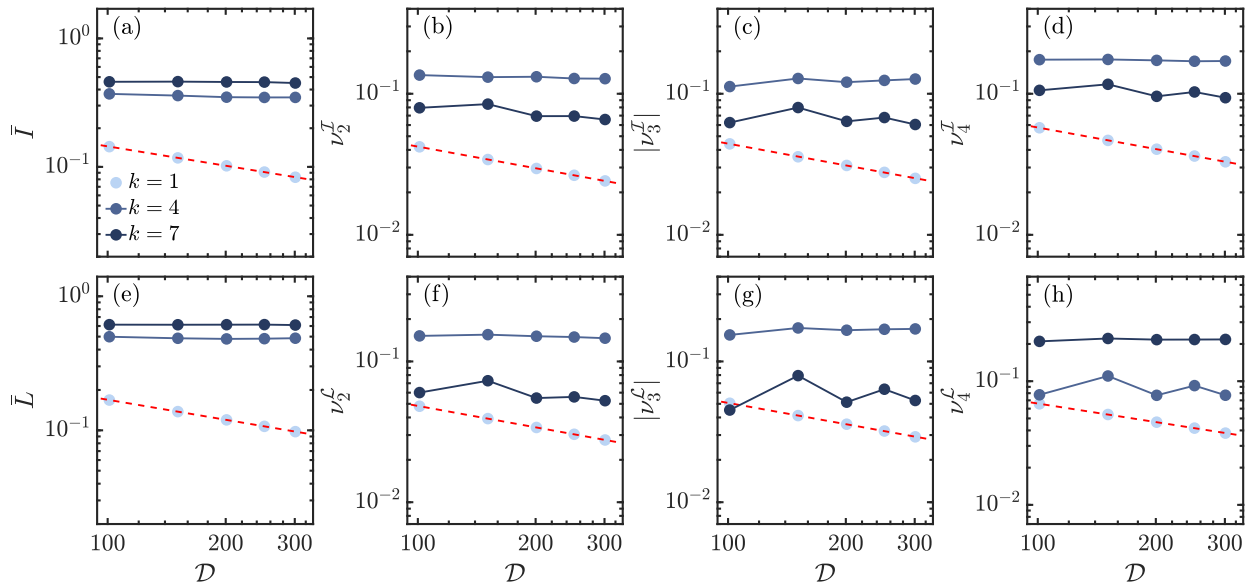}
  \caption{Dependence on the Hilbert space dimension $\mathcal{D}$ of 
  (a) mean value $\bar{I}$, (b) standard deviation $\nu_2^\mathcal{I}$,
  (c) cube root of the third central moment $|\nu_3^\mathcal{I}|$, 
  (d) $\nu_4^\mathcal{I}$ [cf.~Eq.~(\ref{RootMs})] 
  (e) mean value $\bar{L}$, (f) $\nu_2^{\mathcal{L}}$,
  (g) $\nu_3^{\mathcal{L}}$ and (h) $\nu_4^{\mathcal{L}}$ [cf.~Eq.~(\ref{RootMs})] 
  for several kicking strengths $k$, see legend in panel (a). 
  The red dashed line in each panel corresponds to fitting curve of the power law
  $\propto\mathcal{D}^{-\gamma}$ with $\gamma\approx0.5$, to the data of $k=1$ case.
  Other parameter: $\alpha=4\pi/11$.}
  \label{ScLMS}
 \end{figure*}

The visible different scatter plots in Fig.~\ref{LMSEg} imply that 
the localization measures of eigenstates would have different statistical properties 
in the regular and chaotic regimes.
To verify this statement, we study the distribution of the localization measures, denoted by $P(\mathcal{I})$ and $P(\mathcal{L})$, 
which are, respectively, defined as the probability to find $\mathcal{I}_n$ and $\mathcal{L}_n$ in an
infinitesimal interval $[\mathcal{I}, \mathcal{I}+d\mathcal{I}]$ 
and $[\mathcal{L}, \mathcal{L}+d\mathcal{L}]$.

Figure~\ref{PdfLMS} plots $P(\mathcal{I})$ and $P(\mathcal{L})$ for the 
same values of $k$ as in Fig.~\ref{LMSEg}.
One can clearly see that the behaviors of $P(\mathcal{I})$ and $P(\mathcal{L})$ are very similar.
In particular, both of them undergo a drastic change in their property with increasing $k$.
Specifically, in a regular regime with small $k$, the values of $\mathcal{I}_n$ and $\mathcal{L}_n$ 
distribute over a narrow range and are sharply concentrated near $0.1$ and 0.12, respectively.
As a consequence, both $P(\mathcal{I})$ and $P(\mathcal{L})$ 
have a small width and exhibits a sharp peak around 
the upper limit of $\mathcal{I}_n$ and $\mathcal{L}_n$, 
as shown in Figs.~\ref{PdfLMS}(a) and \ref{PdfLMS}(d) [note the scale of the y-axis].
Increasing the kicking strength $k$ tends to increase the width 
of $P(\mathcal{I})$ and $P(\mathcal{L})$, 
as well as shifting the location of the peaks in them to 
larger values of $\mathcal{I}_n$ and $\mathcal{L}_n$.
The distributions of localization measures in the chaotic regime 
are asymmetric or skewed with a peak close to their largest values, 
as evident from Figs.~\ref{PdfLMS}(e) and \ref{PdfLMS}(f).
The asymmetry shape of $P(\mathcal{I})$ and $P(\mathcal{L})$ 
is a consequence of the low-$\mathcal{I}_n/\mathcal{L}_n$ localized eigenstates.

Inspired by our previous works \cite{Batistic2019,Lozej2022,QWang2020}, 
in which the distribution of $\mathcal{L}_n$ for the
delocalized eigenstates in several chaotic systems has been investigated, 
here we explore whether the distributions $P(\mathcal{I})$ and $P(\mathcal{L})$ 
for the delocalized eigenstates can be well described by the beta distribution \cite{Evans2011}
\be \label{BtDs}
   P_\beta(x)=\frac{(x-x_{min})^{a-1}(x_{max}-x)^{b-1}}{(x_{max}-x_{min})^{a+b-1}B(a,b)},
\ee 
where $x_{min}\leq x\leq x_{max}$, $a$ and $b$ are the shape parameters, 
and $B(a,b)=\int_0^1x^{a-1}(1-x)^{b-1}dx$ is the beta function.

The best fitted beta distribution of $P(\mathcal{I})$ is shown in Fig.~\ref{PdfLMS}(c).
The minimal and maximal values of $x$ of our fitted beta distribution 
are empirically found to be $x_{min}=0.145$ and $x_{max}=0.872$, respectively.
Unlike in our previous works, here we fit the distribution on the
interval $[x_{min},x_{max}]$, but note that the actual range of numerically obtained
$\mathcal{I}_n$   is on the interval $[0.1016,0.5975]$. The maximal value
$\mathcal{I}_{n,max}=0.5975$ agrees with the findings in other systems.
Here, the energy is fixed by $j$.
On the other hand, the distribution $P(\mathcal{L})$ is well captured by the beta
distribution fitted on the interval $[x_{min}, x_{max}]=[0.219, 0.8849]$,
as demonstrated in Fig.~\ref{PdfLMS}(f). 
We also note that the data of $\mathcal{L}_n$ are in the range $[0.2479,0.7206]$. 
Thus, the maximal $\mathcal{L}_n$ is found to be $\mathcal{L}_{n,max}=0.7206$, 
which is roughly consistent with other systems.
The fact that the fit by the beta distribution is not perfect should be attributed
to the structure of the chaotic sea in the sense of some stickiness regions
which so far have not been detected.

In fact, more generally, it appears phenomenologically that chaotic quantum eigenstates
exhibit the beta distribution for the localization measures,
provided they are classically uniformly chaotic without significant stickiness regions.
In doing this we consider the eigenstates within a small
energy interval, small enough to have a well defined regime and at the same time large
enough to have a reasonable statistics. This has been demonstrated first in the
mixed-type billiard introduced by Robnik \cite{Batistic2019}, and also in the
stadium billiard of Bunimovich, in the lemon billiards introduced by Heller and Tomsovic
(see Refs. \cite{LLR2021A,Lozej2022} and references therein) and in the Dicke model \cite{QWang2020}.
Therefore, we believe that this distribution of localization measure is universal.
However, if the the chaotic region has significant stickiness
regions such as, e.g., in the ergodic lemon billiard \cite{LLR2021B}, we see
nonuniversal deviations from the beta distribution.

To quantify the above observed features in the distributions of $\mathcal{I}_n$ and $\mathcal{L}_n$,
as well as to quantitatively assess the degree of similarity between them, we consider the mean 
and the $m$th root of the $m$th central moment of $P(\mathcal{O}_q) (q=1,2)$:
\be \label{RootMs}
  \overline{O_q}=\frac{1}{\mathcal{D}}\sum_n\mathcal{O}_{q,n},\ 
  \nu_m^{\mathcal{O}_q}=
  \left[\frac{1}{\mathcal{D}}\sum_n(\mathcal{O}_{q,n}-\overline{O_q})^m\right]^{1/m},
\ee
where $\mathcal{O}_1$ denotes $\mathcal{I}$, $\mathcal{O}_2$ represents $\mathcal{L}$,  
and $\mathcal{D}=j+1$ is the Hilbert space dimension.
As the zeroth central moment is equal to one and the first central moment is zero, 
we are, therefore, mainly interested in $m=2,3,4$, the standard deviation, cube root of the skewness, 
and the fourth root of kurtosis of the distribution, respectively.

Figure~\ref{LMScms} plots these quantities as a function of $k$ for several system sizes $j$.
The onset of chaos can be clearly identified from the sharp growth 
behavior displayed by these quantities.
Remarkably, the drastic growing point of these quantities with increasing $k$ are in agreement with that
of average Lyapunov exponent $\bar{\lambda}$ in Fig.~\ref{PLyE}(d) 
and average gap ratio $\la r\ra$ in Fig.~\ref{Prvsk}(d). 
As $\nu_{2,3,4}^{\mathcal{O}_q}$ quantify the fluctuation, 
the skewness, and the tailedness of the corresponding
distribution, the non-zero values of them in the chaotic regime
imply that the distribution $P(\mathcal{O}_q)$ is asymmetrical, 
consistent with the result shown in Figs.~\ref{PdfLMS}(c) and \ref{PdfLMS}(f). 
Moreover, we also note that our considered quantities are only weakly
dependent on the system size $j$ in the chaotic and regular regime.
%In contrast, for small $k$, we see that these quantities demonstrate an obvious 
%dependence on the system size $j$.
However, on the other hand, they are independent of the kicking strength $k$ in the regular regime.

The dependence of these quantities with varying the 
Hilbert space dimension $\mathcal{D}$ for different system sizes are plotted in Fig.~\ref{ScLMS}.
Notice that we consider the absolute value of $\nu_3^{\mathcal{O}_q}$ rather than 
$\nu_3^{\mathcal{O}_q}$ itself in our numerical simulation.
One can see that all these quantities remain almost unchanged with increasing 
$\mathcal{D}$ for the case with large $k$, wherease they all decrease 
with increasing $\mathcal{D}$ for the case of $k=1$.
Moreover, we find that their decreasing with $\mathcal{D}$
follows the same power law of the form $y=C_y\mathcal{D}^{-\gamma}$ with $\gamma\approx0.5$.
This is explained by the fact that the Husimi function associated with
an invariant torus has roughly length $2\pi$ and thickness
$\approx \sqrt{\hbar_{eff}} = 1/\sqrt{j}$.
This suggests that all of them vanish as $\mathcal{D}\to\infty$ and the distributions 
$P(\mathcal{I})$ and $P(\mathcal{L})$ become the $\delta$-distribution, 
$P(\mathcal{I})=\delta(\mathcal{I}), P(\mathcal{L})=\delta(\mathcal{L})$, as expected.
It should be observed that according to the Figs. \ref{LMScms} and \ref{ScLMS}
in the chaotic regime the average value of $\mathcal{O}_q$ 
has converged to its semiclassical limit $j\rightarrow \infty$,
while $\nu_2^{\mathcal{O}q}$, the standard deviation, seems to decay very slowly 
with $j$ to its expected semiclassical limit  $\nu_2^{\mathcal{O}_q}=0$. The case
$k=4$ shows no decay with $\mathcal{D}=j+1$, probably because it is still a mixture of
a large chaotic component and a small regular component.

 \begin{figure*}
  \includegraphics[width=\textwidth]{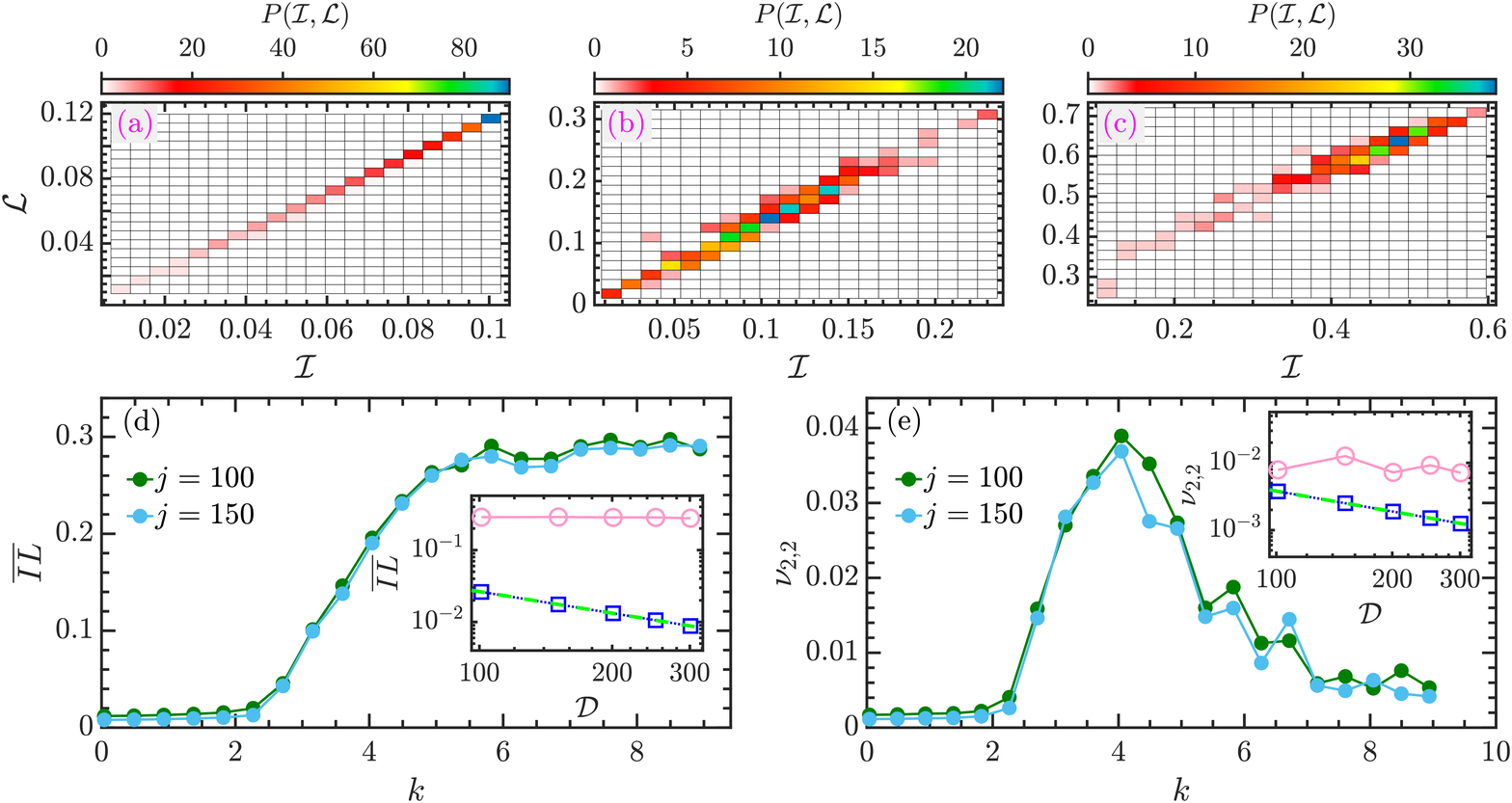}
  \caption{Joint probability distribution $P(\mathcal{I},\mathcal{L})$ of the kicked top model
  for $k=0.4$ (a), $k=2.4$ (b), and $k=6$ (c) with $j=150$.
  (d) Dependence of $\overline{IL}$ [cf.~Eq.~(\ref{CMJoint})] on the kicking strength $k$
  for different system sizes $j$. 
  Inset: Scaling of $\overline{IL}$ with Hilbert space dimension $\mathcal{D}$ for $k=1$ (blue squares)
  and $k=7$ (pink circles). The green dashed line in the inset marks the scaling 
  $\overline{IL}\sim\mathcal{D}^{-1}$.
  (e) Evolution of $\nu_{2,2}$ [cf.~Eq.~(\ref{CMJoint})] versus 
  the kicking strength $k$ for several system sizes $j$.
  The inset plots $\nu_{2,2}$ as a function of $\mathcal{D}$ for $k=1$ (blue squares) and 
  $k=7$ (pink circles). In the inset, the green dashed line denotes the decay 
  $\nu_{2,2}\sim\mathcal{D}^{-\zeta}$ with $\zeta=1$. 
   Other parameter: $\alpha=4\pi/11$.}
  \label{JointPdf}
\end{figure*}

\subsection{Joint probability distribution of the localization measures}

We finally discuss the interplay between the onset of chaos and the properties 
of the joint probability distribution of the localization measures. 
The above revealed statistical properties of the individual localization measure distributions indicate
that the statistics of the localization measures provides useful information about the structure 
of the eigenstates and can detect the onset of chaos.
Further insights into the characteristics of the eigenstates and the signatures of quantum chaos can be
obtained from the joint probability distribution $P(\mathcal{I},\mathcal{L})$, 
which is defined as the probability to find $\mathcal{I}$ and $\mathcal{L}$ in an infinitesimal
box $[\mathcal{I},\mathcal{I}+d\mathcal{I}]\times[\mathcal{L},\mathcal{L}+d\mathcal{L}]$.

The joint distribution $P(\mathcal{I},\mathcal{L})$ of the kicked top model for various values of 
the kicking strength $k$ are shown in Figs.~\ref{JointPdf}(a)-\ref{JointPdf}(c).
For the regular case with $k=1$, the distribution $P(\mathcal{I},\mathcal{L})$ distributes over
the small values of $\mathcal{I}$ and $\mathcal{L}$ with a very narrow width 
in the $\mathcal{I}\mathcal{L}$ plane [Fig.~\ref{JointPdf}(a)]. 
As the kicking strength $k$ increases, the joint distribution extends out to large $\mathcal{I}$ and 
$\mathcal{L}$.
Moreover, the width of the distribution $P(\mathcal{I},\mathcal{L})$ also increases with increasing
$k$, as seen in Figs.~\ref{JointPdf}(b) and \ref{JointPdf}(c).
However, we note that the largest width of $P(\mathcal{I},\mathcal{L})$ occurs in the mixed regime,
instead of the fully chaotic case.
In fact, most of the eigenstates are delocalized in strong chaotic regime, both
$\mathcal{I}$ and $\mathcal{L}$ are concentrated around some large values, 
resulting in small width of the joint distribution $P(\mathcal{I},\mathcal{L})$.
It has been found already in Ref. \cite{Batistic2020} that the two localization measures,
after a proper normalization, are nearly linearly related.

The observed characters of $P(\mathcal{I},\mathcal{L})$ can be quantified by 
the following mixed moments, 
\be \label{CMJoint}
  \overline{IL}=\frac{1}{\mathcal{D}}\sum_{n}\mathcal{I}_n\mathcal{L}_n,\ 
  \nu_{2,2}=\left[\frac{1}{\mathcal{D}}\sum_n(\mathcal{I}_n-\bar{I})
     (\mathcal{L}_n-\bar{L})\right]^{1/2},
   \ee
where $\mathcal{D}=j+1$ is the Hilbert space dimension, and $\bar{I}$ and $\bar{L}$ are the mean
values of $\mathcal{I}_n$ and $\mathcal{L}_n$ in Eqs.~(\ref{RootMs}).
The evolution of these quantities as a function of $k$ is depicted in Figs.~\ref{JointPdf}(d) and
\ref{JointPdf}(e).
Clearly, we see that the transition to chaos also leaves an imprint in the joint distibution 
of the localization measures and the onset of chaos with increasing 
kicking strength can be unveiled unambiguously by the upturn of these quantities as $k$ increases. 
Furthermore, we also find that the values of these quantities are almost independent of the system
size for the chaotic case, while in the regular regime they decrease with increasing 
system size and are decaying as $\sim\mathcal{D}^{-\zeta}$ with $\zeta\approx 1$,
as demonstrated in the insets of Figs.~\ref{JointPdf}(d) and \ref{JointPdf}(e).
Again, this is due to the expected proportionality of these quantities
to the effective Planck constant $\hbar_{eff}  \approx 1/j$. 
Hence, in the regular regime, the joint distribution approaches 
the Dirac $\delta$-distribution as $j\to\infty$, in agreement with the asymptotic 
behaviors of $P(\mathcal{I})$ and $P(\mathcal{L})$.
Here, again we may emphasize that the average value $\overline{IL}$ 
in the chaotic regime has converged to its semiclassical value as $j\rightarrow \infty$, while
the standard deviation $\nu_{2,2}$ decays very slowly with $j$.

\section{Conclusion} \label{summary}

 In this work, the phase space localization properties of the eigenstates have been
 scrutinized in the kicked top model that undergoes a transition to chaos 
 with increasing the kicking strength. 
 The information about the phase space localization of the eigenstates is encoded in the 
 associated Husimi functions, which exhibit distinct localization features depending on whether
 the system is regular or chaotic.
 Hence, the emergence of chaos bears a significant change in the eigenstates and the  
 crossover from integrability/regularity to quantum chaos can be probed by
 the localization characters of eigenstates in phase space. 
 Remarkably, we have again found that there still exist some localized eigenstates 
 even in the deep chaotic regime, which is a manifestation of quantum
 dynamical localization, and the localization measures exhibit the
 beta distribution.

 The notably different localization behavior of the Husimi function in regular and chaotic regime 
 has led to a characterization of the phase space localization phenomenon of eigenstates in terms of 
 two different localization measures, 
 i.~e., the inverse participation ratio and the Wehrl entropy, 
 that are based on the Husimi function. 
 The investigation of the statistics of the localization measures reveals that their 
 distributions are sharply peaked around some small values in the regular regime, indicating
 that the eigenstates are highly localized states in the phase space, spanned by the
 invariant tori in the classical phase space.
 As the system tends from integrability/regularity toward chaos, 
 the width of their distributions is increased and the peak of the distributions is moved to the large
 values of the localization measures. The scenario is in line with the predictions
 of the Principle of Uniform Semiclassical Condensation (PUSC) of the Husimi (or Wigner) functions,
 in the semiclassical limit.
 However, we have found that the distribution of the localization measures also
 displays fluctuations in the deep chaotic regime. In fact, the distribution of the localization
 measure approaches the beta distribution, in agreement with previous works \cite{Lozej2022},
 and in the ultimate semiclassical limit is expected to tend to the Dirac delta distribution.
 In this ultimate limit (not yet seen in this study)
 most of eigenstates of the fully chaotic system are then expected to be uniformly delocalized in the 
 phase space. 
 Thus, the features shown by the distributions of the localization measures certainly
 confirm their usefulness to detect the onset of chaos.
 
 To capture the quantitative features of the distributions of the localization measures,
 we also consider the central moments of the distributions.  
 We have demonstrated that the central moments are sensitive to the presence of chaos,
 which results in a drastic change in the behaviors of the central moments.
 We therefore verified that the statistical properties of the localization measures 
 are the useful witnesses of chaos.
 In particular, as we showed, the transition to chaos provided by the central moments is in good
 agreement with the classical case. 
 Further analysis on the scaling of the central moments reveals that the distribution of the
 localization measures in the regular regime is approaching the Dirac delta distribution
 in the classical limit, as expected, while in the fully chaotic regime in the same limit
 it approaches the delta distribution peaked at the maximal value of the localization measure.
 This approach seems to be extremely slow.

 The results presented in this work provide more insights into the relationship 
 between the phase space structure of eigenstates and the onset of chaos in quantum systems.
 Unveiling how the statistics of the localization measures of eigenstates is affected by
 underlying chaos would help us get deep understanding on the signatures of quantum chaos
 and opens up a new way to distinguish between regular and chaotic dynamics in quantum systems.
 An interesting extension of this work is to explore how our results change in the many body quantum
 chaotic systems, such as the coupled top model \cite{Mondal2020}, 
 Dicke model \cite{QWang2020,Villasenor2021,Pilatowsky2022}, 
 and Bose-Hubbard model \cite{Pausch2021, Fogarty2021,Wittmann2022}.
 Another open question that deserves examination is to study the correlation between 
 the phase space  localization measures and the entanglement entropy in quantum chaotic systems.

 \acknowledgements

 This work was supported by the Slovenian Research Agency (ARRS) under Grant 
 No.~J1-4387. 
 Q.~W. acknowledges support from the National Science Foundation of China under 
 Grant No.~11805165, and
 Zhejiang Provincial Nature Science Foundation under Grant No.~LY20A050001.

 \section*{Appendix: A short derivation of the formulae (\ref{QEJ})}

 The discrete mapping of the the Heisenberg operators $\mathbf{J}=(J_x,J_y,J_z)$ is a composition of two mappings:
 the first one is a free precessional rotation (between two torsional kicks) by the angle $\alpha$ around the $z$ axis,
 namely  $\bar{J_\ell} = \exp (i\alpha J_z) \; J_\ell \; \exp (-i\alpha J_z)$ with $\ell=x, y, z$. 
 Using the Campbell identity [Eq.~(\ref{Campbell})] and
 the commutation relations for $\mathbf{J}=(J_x,J_y,J_z)$ we find immediately that
 \be  \label{freeprecession}
 \bar{J_z} = J_z, \;\;\; (\bar{J}_x+i\bar{J}_y)= (J_x+iJ_y) \exp (i\alpha).
\ee
The other mapping is due to a torsional kick, which appears periodically with period one, generated by
$\bar{J_\ell} = \exp (i\omega J_x^2)\; J_\ell \;\exp (-i\omega J_x^2)$, where $\omega = k/(2j)$. 
When applying again the Campbell
identity, we first note that, using the commutation relations for $\mathbf{J}=(J_x,J_y,J_z)$ , the following commutator
is found to be $[J_x^2, (J_y+iJ_z)(2J_x+1)^n] = (J_y+iJ_z)(2J_x+1)^{(n+1)}$, for all nonnegative  $n=0,1,2,\dots$, and therefore,
\be  \label{kick}
 \bar{J_x} = J_x, \;\;\; (\bar{J}_y+i\bar{J}_z)= (J_y+iJ_z) \exp (i\omega (2J_x+1)).
\ee
Compositum of the two mappings results in the transformation in Eq. (\ref{QEJ}).

\bibliographystyle{apsrev4-1}
\bibliography{LocKT}

\end{document}